\documentclass[aps,pra,preprint,superscriptaddress,showpacs,showkeys,nofootinbib]{revtex4-1}
\usepackage{CJK}
\usepackage{amsfonts}
\usepackage{graphicx}
\usepackage{amssymb}
\usepackage{amsmath}
\usepackage{natbib}

\newcommand{\ket}[1]{\left | #1 \right \rangle}
\newcommand{\bra}[1]{\left \langle #1 \right |}

\begin{document}
%\begin{CJK*}{GBK}{}
\title{Impact of imperfect measurements on multi-party quantum key distribution}

\author{Yang Xiang}
\email{xiangyang@vip.henu.edu.cn (corresponding author)}
\affiliation{School of Physics and Electronics, Henan University, Kaifeng, Henan 475004, China}

\date{\today}
\begin{abstract}

The imperfection of measurements in real-world scenarios can compromise the performance of device-independent quantum key distribution (DIQKD) protocols.
In this paper, we focus on a specific DIQKD protocol that relies on the violation of the Svetlichny's inequality (SI), considering an eavesdropper utilizing the convex combination attack. Our analysis covers both the three-party DIQKD case and the general $n$-party scenario. Our main result is the relationship between the measurement accuracy and the extractable secret-key rate in all multi-party scenarios. The result demonstrates that as measurement accuracy improves, the extractable secret-key rate approaches $1$, reaching its maximum value when the measurement accuracy is perfect. We reveal that achieving positive extractable secret-key rates requires a threshold measurement accuracy that is consistently higher than the critical measurement accuracy necessary to violate the SI. We depict these thresholds for $n$-party scenarios ranging from $n=3$ to $n=10$, demonstrating that as the number of parties ($n$) increases, both thresholds exhibit a rapid and monotonic convergence towards unity. Furthermore, we consider a scenario involving a non-maximally entangled state with imperfect measurements, where the emission of the initial GHZ state undergoes noise during transmission, resulting in a Werner state. The study further quantifies and demonstrates the relationship between the extractable secret-key rate, the visibility of the Werner state, and the measurement accuracy, specifically emphasizing the three-party scenario. This study aims to illuminate the influence of imperfect measurement accuracy on the security and performance of multi-party DIQKD protocols. The results emphasize the importance of high measurement accuracy in achieving positive secret-key rates and maintaining the violation of the SI.

\end{abstract}
%\keywords{multipartite quantum key distribution, genuine multipartite quantum correlations, Svetlichny's inequality,
%device-independent quantum key distribution, measurement accuracy}

\maketitle
%\end{CJK*}

% insert suggested keywords - APS authors don't need to do this
%\keywords{}

%\maketitle must follow title, authors, abstract, \pacs, and \keywords

% body of paper here - Use proper section commands
% References should be done using the \cite, \ref, and \label commands

\section{Introduction}

 Quantum Key Distribution (QKD) is considered a groundbreaking technology that utilizes the principles of quantum mechanics to establish a secure key for encrypting and decrypting information. Unlike traditional encryption methods, QKD offers genuine security by relying on the fundamental properties of quantum mechanics \cite{bennett1984quantum}. However, traditional QKD protocols often rely on assumptions about the security of devices, which can introduce potential vulnerabilities. To address these concerns, device-independent quantum key distribution (DIQKD) has emerged as a promising approach \cite{PhysRevLett.67.661,bennett1992quantum,PhysRevLett.98.230501,Pironio_2009}. DIQKD aims to provide security guarantees without relying on assumptions about the internal workings of devices, but instead utilizes the principles of quantum entanglement and Bell inequalities \cite{bell1964einstein,bell1988speakable,PhysRevLett.23.880} to ensure the security of key distribution.
When it comes to two-party communication, there exists a wide range of QKD protocols \cite{PhysRevLett.68.557,RevModPhys.74.145,gisin1997quantum,PhysRevLett.87.117901,PhysRevA.65.012311,PhysRevLett.86.1911,PhysRevLett.87.010403,
PhysRevLett.90.160408,beige2002secure,PhysRevA.67.012311}. Some of these protocols have even found applications in commercial settings \cite{PhysRevLett.119.200501,pirandola2020advances,PRXQuantum.2.010304,PRXQuantum.3.020341}. QKD is one of the fastest-growing areas in the field of quantum information \cite{PhysRevLett.94.230504,RevModPhys.81.1301,PhysRevA.93.032338,
PhysRevA.93.052303,PhysRevA.95.010101,PhysRevLett.118.220501,PhysRevA.100.042329}.

 Afterward, to securely distribute multi-party keys and accomplish tasks that two-party keys cannot handle, multipartite QKD, also known as quantum conference key agreement(QCKA), has emerged as a prominent research area in recent years \cite{Grasselli2018,Grasselli2019,PhysRevLett.114.090501,sciadv.abe0395,PhysRevLett.108.100402,
PhysRevA.97.022307,PhysRevResearch.2.023251,yinieee2021,Caoyin2021,Liyin2021,Fan-Yuan22,wang2022twin}. In the secure distribution of multi-party keys, the presence of genuine multipartite quantum correlations (GMQC) \cite{PhysRevLett.106.020405,PhysRevLett.106.250404,PhysRevLett.88.210401,PhysRevLett.107.210403,PhysRevA.89.032117,RevModPhys.86.419} is required. GMQC refers to collective correlations involving all subsystems that cannot be expressed as mixtures of states where only a subset of subsystems are entangled. There exist various inequivalent types of GMQC \cite{xiang2011bound}, and its structure is significantly more intricate than that of bipartite correlations \cite{PhysRevA.71.022101,pironio2011extremal}. Svetlichny introduced the first method to detect GMQC, known as Svetlichny's inequality (SI) \cite{PhysRevD.35.3066,PhysRevLett.89.060401}, which serves as a Bell-like inequality. Violation of the SI confirms the presence of GMQC. GHZ states, a kind of quantum state exhibiting GMQC \cite{greenberger1990bell,pan2000experimental}, can be employed along with specific measurement settings to achieve the maximum violation of the SI \cite{PhysRevLett.89.060401}. Several multi-party DIQKD protocols have already been developed, where their security is guaranteed by violations of different quantum nonlocality inequalities \cite{PhysRevLett.108.100402,PhysRevA.97.022307,PhysRevResearch.2.023251}. Recently, we proposed a DIQKD protocol for secure multi-party key distribution, employing the violation of SI as the security criterion \cite{Xiang2023}. Our multipartite DIQKD protocol distinguishes itself from the conventional QCKA. While in the conventional QCKA, every party shares the same key, our multipartite DIQKD protocol features keys that are collectively correlated among the involved parties, rather than being pairwise correlated. This unique property empowers our multipartite key to not only function as a two-party key but also to surpass the limitations of two-party keys and enable various tasks that were previously unattainable \cite{Xiang2023}.

While DIQKD has demonstrated significant potential, the performance of DIQKD protocols is influenced by various factors. This is primarily due to the fact that the security of existing DIQKD protocols relies on the violation of Bell inequalities, making them susceptible to attacks through experimental loopholes. Two main types of experimental loopholes exist: the locality loophole and the detection loophole. The locality loophole is not a major issue in DIQKD since it assumes that the users' locations are secure, preventing unwanted information leakage. However, the detection loophole poses a more significant challenge. In practical DIQKD processes, there may be losses in the transmission of emitted quantum bits, and the detector efficiency is not perfect, leading to the possibility of undetected quantum bits and false alarms. As a result, the detection loophole may enable eavesdroppers to intervene during the detection process, compromising the security of the DIQKD system. This issue has been extensively studied in the context of two-party DIQKD. For traditional DIQKD protocols \cite{Pironio_2009,masanes2011,reichardt2013,PhysRevLett.113.140501,arnon2018}, high detector efficiency (e.g., $\eta>90\%$) is required. In recent years, various approaches for two-party DIQKD have been proposed, aiming to lower the threshold efficiency \cite{PhysRevLett.124.230502,Woodhead2021deviceindependent,
Sekatski2021deviceindependent,PhysRevA.103.052436,PhysRevLett.128.110506,Schwonnek2021,PhysRevLett.124.020502,brown2021}.

To the best of our knowledge, there is still a lack of research on the influence of the detection loophole on multi-party DIQKD. The objective of this paper is to investigate the effects of imperfect measurements on multi-party DIQKD and specifically explore the correlation between measurement accuracy and the extractable secret-key rate in multi-party scenarios.
In the context of the DIQKD process, measurement accuracy comprises two critical aspects: (1) the detection efficiency of the measuring instrument in correctly identifying transmitted quantum bits, and (2) the precision of reflecting the state of the transmitted quantum bits by the measuring instrument. The latter, which refers to the accurate rate of state reflection, is primarily determined by the overall performance of the measuring instrument. On the other hand, the detection efficiency is influenced by the losses that occur during the transmission of emitted quantum bits. To establish a quantitative relationship among the accurate rate of state reflection (denoted as "$q_{1}$"), the detection efficiency (denoted as "$q_{2}$"), and the measurement accuracy (denoted as "$p$"), we need to decide how to deal with all undetected events. Currently, two main approaches are commonly employed to handle undetected events. The first approach is to completely disregard these undetected events and assume that the detected events adequately represent all events (based on the fair sampling assumption). In this approach, it is apparent that $p=q_{1}$. The second approach addressing imperfect detection efficiency assign all undetected events to the same measurement result\cite{Pironio_2009,PhysRevLett.128.110506,PhysRevA.103.052436}. If we assume that the expected measurement outcomes ($0$ or $1$) are unbiased, we can establish the quantitative relationship in the second approach as follows: $p=q_{1}q_{2}+\frac{1-q_{2}}{2}$ . So measurement accuracy refers to the degree to which the measurement results align with the theoretical or expected values, encompassing the detection efficiency as a broader concept.

In this study, we investigate the impact of imperfect measurement accuracy on our multi-party DIQKD protocol \cite{Xiang2023}, which relies on the violation of the SI, assuming an eavesdropper employing the convex combination attack \cite{Acin2006,PhysRevLett.97.120405,PhysRevLett.127.050503}. Our main result of this paper is the relationship between the measurement accuracy and the extractable secret-key rate in both the three-party case and the general multi-party scenario. Our findings demonstrate that as the measurement accuracy improves, the extractable secret-key rate steadily approaches $1$. We also make the following observations: in all scenarios, the threshold measurement accuracy required to achieve positive extractable secret-key rates is always higher than the critical measurement accuracy needed to violate the corresponding SI. We present a graphical representation of these thresholds for $n$-party scenarios ranging from $n=3$ to $n=10$. The graph demonstrates that as the value of $n$ increases, both thresholds show a trend of approaching unity monotonically and rapidly. Furthermore, we consider a combined scenario involving a non-maximally entangled state with imperfect measurement. We assume that the initial GHZ state emitted is subject to noise during transmission, resulting in a Werner state \cite{PhysRevA.40.4277}. In this context, we calculate and demonstrate the relationship between the extractable secret-key rate, the visibility of the Werner state, and the measurement accuracy, specifically focusing on the three-party scenario. Overall, this study provides insights into the influence of imperfect measurement accuracy on the security and performance of multi-party DIQKD protocols, highlighting the importance of high measurement accuracy to achieve positive secret-key rates and maintain the violation of the SI.

%%%%%%%%%%%%%%%%%%%%%%%%%%%%%%%%%%%%%%%%%%%%%%%%%%%%%%%%%%%%%%%%%%%%%%%%%%%%%%%%%%%%%%%%%%%%%%%%%%%%%%%%%%%%%%%%%%%

%%%%%%%%%%%%%%%%%%%%%%%%%%%%%%%%%%%%%%%%%%%%%%%%%%%%%%%

\section{Three-party key distribution with imperfect measurement accuracy}

\subsection{The critical measurement accuracy for the violation of SI}
We will begin by providing a brief introduction to our three-party DIQKD protocol \cite{Xiang2023}. Alice, Bob, and Carol generate their individual three-party keys by measuring the GHZ state $\ket{\Psi}_{GHZ}=\frac{1}{\sqrt{2}}\left(\ket{\uparrow\uparrow\uparrow}+\ket{\downarrow\downarrow\downarrow}\right)$, while ensuring security through the violation of the SI. In this paper, we use $\uparrow$ ($\downarrow$) to denote spin up (spin down) along the $z$-axis direction.
We denote the spin measurements of the three individuals as $\hat{A}_{i}$, $\hat{B}_{j}$, and $\hat{C}_{k}$ respectively. We require that all spin measurements lie within the $x-y$ plane, which allows us to represent these measurements using their azimuthal angles $\alpha_{i}$, $\beta_{j}$, and $\gamma_{k}$ as follows: $\hat{A}_{i}=\cos{\alpha_{i}}\sigma_{x}+\sin{\alpha_{i}}\sigma_{y}$, $\hat{B}_{j}=\cos{\beta_{j}}\sigma_{x}+\sin{\beta_{j}}\sigma_{y}$ , and $\hat{C}_{k}=\cos{\gamma_{k}}\sigma_{x}+\sin{\gamma_{k}}\sigma_{y}$.
For simplicity, we will represent the measurement settings as $(\alpha_{i}, \beta_{j}, \gamma_{k})$. Since $\ket{\Psi}_{GHZ}$ is the common eigenstate of $\sigma_{x}\otimes\sigma_{y}\otimes\sigma_{y}$, $\sigma_{y}\otimes\sigma_{x}\otimes\sigma_{y}$, $\sigma_{y}\otimes\sigma_{y}\otimes\sigma_{x}$, and $\sigma_{x}\otimes\sigma_{x}\otimes\sigma_{x}$ with corresponding eigenvalues of $-1$, $-1$, $-1$, and $1$ respectively, Alice, Bob, and Carol can extract raw keys from the outcomes of the measurements corresponding to the settings $(0,0,0)$, $(0,\frac{\pi}{2},\frac{\pi}{2})$, $(\frac{\pi}{2},0,\frac{\pi}{2})$, and $(\frac{\pi}{2},\frac{\pi}{2},0)$. Additionally, by utilizing the measurement protocol $\{\alpha_{0}=-\frac{\pi}{4}, \alpha_{1}=\frac{\pi}{4}, \beta_{0}=0, \beta_{1}=\frac{\pi}{2}, \gamma_{0}=0, \gamma_{1}=\frac{\pi}{2}$\} in conjunction with the GHZ state, the maximum quantum violation $4\sqrt{2}$ of the following SI Eq. (\ref{si}) can be achieved,
\begin{eqnarray}
&&\big|\langle A_{0}B_{0}C_{0}\rangle+\langle A_{0}B_{0}C_{1}\rangle+\langle A_{0}B_{1}C_{0}\rangle+\langle A_{1}B_{0}C_{0}\rangle\nonumber\\
&&-\langle A_{0}B_{1}C_{1}\rangle-\langle A_{1}B_{0}C_{1}\rangle-\langle A_{1}B_{1}C_{0}\rangle-\langle A_{1}B_{1}C_{1}\rangle\big|\leq 4.
\label{si}
\end{eqnarray}
Therefore, Alice, Bob, and Carol can utilize the outcomes of the measurements corresponding to the settings $(\pm\frac{\pi}{4},0,0)$, $(\pm\frac{\pi}{4},0,\frac{\pi}{2})$, $(\pm\frac{\pi}{4},\frac{\pi}{2},0)$, and $(\pm\frac{\pi}{4},\frac{\pi}{2},\frac{\pi}{2})$ to calculate the SI value and confirm the security of the DIQKD protocol.
In our protocol, Alice randomly choose to perform a measurement from  \{${\alpha_{0}=-\frac{\pi}{4}, \alpha_{1}=\frac{\pi}{4}, \alpha_{2}=0,\alpha_{3}=\frac{\pi}{2}}$\}, Bob randomly choose a measurement from \{${\beta_{0}=0, \beta_{1}=\frac{\pi}{2}}$\}, and similarly Carol randomly choose a measurement from \{${\gamma_{0}=0, \gamma_{1}=\frac{\pi}{2}}$\}. They then use the results of the aforementioned measurement settings to extract the raw key and confirm the violation of the SI, respectively. For more detailed information, please refer to \cite{Xiang2023}.

To facilitate the study of measurement accuracy, we consider the SI in the form of probabilities:
\begin{eqnarray}
\frac{1}{8}\sum_{x,y,z}P(a\oplus b\oplus c=xy\oplus yz\oplus xz|x,y,z)\leq \frac{3}{4}.
\label{si2}
\end{eqnarray}
In this formulation, the variables $x$, $y$, and $z$ take on values of $0$ or $1$, representing the measurement choices of Alice, Bob, and Carol, respectively. For instance, when $x=0$, it indicates that Alice chooses measurement $\hat{A}_{0}$, whereas $x=1$ represents Alice's choice of measurement $\hat{A}_{1}$.
The variables $a$, $b$, and $c$ correspond to the measurement outcomes and can each take on values of $0$ or $1$. The relationship between these variables and the measurement values $A_{x}$, $B_{y}$, and $C_{z}$ in Eq. (\ref{si}) is defined as follows: $a=\frac{1-A_{x}}{2}$, $b=\frac{1-B_{y}}{2}$, and $c=\frac{1-C_{z}}{2}$. In Eq.(\ref{si2}), the probabilities are conditional probabilities. For example, $P(a\oplus b\oplus c=xy\oplus yz\oplus xz|x=0,y=0,z=0)$ is the probability that $a\oplus b\oplus c=xy\oplus yz\oplus xz$ under the condition $x=0,y=0,z=0$. By considering these relationships, we can easily derive Eq. (\ref{si2}) from Eq. (\ref{si}). The quantum bound for Eq. (\ref{si2}) is $(\frac{1}{2}+\frac{\sqrt{2}}{4})$, which corresponds to the quantum bound of $4\sqrt{2}$ in Eq. (\ref{si}).

Without loss of generality, this study focuses on isotropic correlations, where the probability $P(a\oplus b\oplus c=xy\oplus yz\oplus xz|x,y,z)$ remains the same for different measurement choices $(x, y, z)$. In \cite{Xiang2012}, it was proven that any multipartite no-signaling correlation can be transformed into an isotropic form using a local randomization procedure. Thus, for the purposes of this analysis, we consider correlations with this desirable isotropic property.
Furthermore, we introduce the concept of measurement accuracy represented by $p$. This parameter denotes the probability that the measuring instrument successfully detects and accurately reflects the state of the transmitted quantum bits. Importantly, we assume that the measurement accuracy $p$ is independent of the specific measurement selections made by the parties involved. This means that the accuracy remains consistent regardless of the particular measurements chosen.
By adopting these assumptions and considerations, we can investigate the influence of measurement accuracy on the inequality as defined in Eq. (\ref{si2}).

We assume that in the case of perfect measurements, the probability distribution of the measurement results is represented by $P^{\prime}(a\oplus b\oplus c=xy\oplus yz\oplus xz|x,y,z)$. If we consider a measurement accuracy of $p$, the probability distribution of the measurement results will be denoted as $P(a\oplus b\oplus c=xy\oplus yz\oplus xz|x,y,z)$. Our aim now is to establish the relationship between these two probability distributions. We begin by examining the scenario of perfect measurements. If the total number of events (i.e., measurement results) corresponding to the measurement setting $(x, y, z)$ is $M^{\prime}_{1}+M^{\prime}_{2}$, where $M^{\prime}_{1}$ represents the number of events that satisfy the condition $a\oplus b\oplus c=xy\oplus yz\oplus xz$, and $M^{\prime}_{2}$ represents the number of events that do not satisfy the condition, then we can express this relationship as:
\begin{eqnarray}
\frac{M^{\prime}_{1}}{M^{\prime}_{1}+M^{\prime}_{2}}=P^{\prime}(a\oplus b\oplus c=xy\oplus yz\oplus xz|x,y,z).
\label{p1}
\end{eqnarray}
Next, let's consider the scenario where the measurement accuracy is denoted as $p$. Assuming that the total number of events corresponding to the measurement setting $(x, y, z)$ is $M_{1}+M_{2}$, where $M_{1}$ represents the number of events that satisfy the condition $a\oplus b\oplus c=xy\oplus yz\oplus xz$, and $M_{2}$ represents the number of events that do not satisfy the condition, we can determine a quantitative relationship between $M^{\prime}_{1}$, $M^{\prime}_{2}$, $M_{1}$, and $M_{2}$ as follows:
\begin{eqnarray}
M_{1}=p^3 M^{\prime}_{1}+3p(1-p)^2 M^{\prime}_{1}+3p^2(1-p) M^{\prime}_{2}+(1-p)^3 M^{\prime}_{2}\nonumber\\
M_{2}=p^3 M^{\prime}_{2}+3p(1-p)^2 M^{\prime}_{2}+3p^2(1-p) M^{\prime}_{1}+(1-p)^3 M^{\prime}_{1}.
\label{mm}
\end{eqnarray}
Then, we can derive the probability distribution $P(a\oplus b\oplus c=xy\oplus yz\oplus xz|x,y,z)$ for the scenario where the measurement
accuracy is $p$:
\begin{eqnarray}
\frac{M_{1}}{M_{1}+M_{2}}&=&P(a\oplus b\oplus c=xy\oplus yz\oplus xz|x,y,z)\nonumber\\
&=&\big[p^3+3p(1-p)^2\big]\frac{M^{\prime}_{1}}{M^{\prime}_{1}+M^{\prime}_{2}}
+\big[3p^2(1-p)+(1-p)^3\big]\frac{M^{\prime}_{2}}{M^{\prime}_{1}+M^{\prime}_{2}}\nonumber\\
&=&\big[p^3+3p(1-p)^2\big]P^{\prime}(a\oplus b\oplus c=xy\oplus yz\oplus xz|x,y,z)\nonumber\\
&&+\big[3p^2(1-p)+(1-p)^3\big]\big[1-P^{\prime}(a\oplus b\oplus c=xy\oplus yz\oplus xz|x,y,z)\big]\nonumber\\
&=&(2p-1)^3 P^{\prime}(a\oplus b\oplus c=xy\oplus yz\oplus xz|x,y,z)+\frac{1-(2p-1)^3}{2}.
\label{p2}
\end{eqnarray}
From the equation above, we can derive:
\begin{eqnarray}
&&\frac{1}{8}\sum_{x,y,z}P(a\oplus b\oplus c=xy\oplus yz\oplus xz|x,y,z)\nonumber\\
&=&(2p-1)^3 \left[\frac{1}{8}\sum_{x,y,z}P^{\prime}(a\oplus b\oplus c=xy\oplus yz\oplus xz|x,y,z)\right]+\frac{1-(2p-1)^3}{2}.
\label{si3}
\end{eqnarray}
In the scenario where measurements are perfect, the maximum quantum violation of the SI can be achieved for the GHZ state with a suitable measurement protocol. This implies that we can obtain a value of $\frac{1}{8}\sum_{x,y,z}P^{\prime}(a\oplus b\oplus c=xy\oplus yz\oplus xz|x,y,z)=\frac{1}{2}+\frac{\sqrt{2}}{4}$ at most. Therefore, from the equation Eq. (\ref{si3}), we can conclude that only when $p>\frac{1}{2}(1+\frac{1}{2^{1/6}})$ ($\approx 0.945449$), the probability distribution $P(a\oplus b\oplus c=xy\oplus yz\oplus xz|x,y,z)$ can result in a violation of SI. We refer to this critical measurement accuracy as $p_{cr}$.

\subsection{The threshold measurement accuracy for the positive extractable secret-key rates}

We consider the scenario where the eavesdropper, Eve, employs a convex combination attack. In the context of DIQKD, we assume that Eve has control over the particle source, and the three legitimate users publicly announce their measurement choices for each round. This allows Eve to mimic the correlation observed by the legitimate users, which can be expressed by the following equation
\begin{eqnarray}
P(a\oplus b\oplus c=xy\oplus yz\oplus xz|x,y,z)&=&q_{L} P^{L}(a\oplus b\oplus c=xy\oplus yz\oplus xz|x,y,z)\nonumber\\
&+&(1-q_{L})P^{NL}(a\oplus b\oplus c=xy\oplus yz\oplus xz|x,y,z).
\label{cca}
\end{eqnarray}
Where $q_{L}\in[0,1]$, we refer to it as the local weight. The term $P^{L}(a\oplus b\oplus c=xy\oplus yz\oplus xz|x,y,z)$ represents the local correlation, which means it is Svetlichny local and does not violate SI. On the other hand, $P^{NL}(a\oplus b\oplus c=xy\oplus yz\oplus xz|x,y,z)$ denotes a carefully chosen nonlocal quantum correlation that can result in the violation of SI.
From Equation  Eq. (\ref{cca}), it can be observed that in order to mimic the correlation observed by the legitimate users, Eve distributes local deterministic correlations with a probability of $q_{L}$, while distributing nonlocal quantum correlations with a probability of $1-q_{L}$. In cases where Eve distributes local deterministic correlations, she has complete information about the outcomes of the legitimate users' measurements. Therefore, it is in her best interest to maximize the value of $q_{L}$. This attack strategy is known as the convex combination attack.

The correlation strength of $P(a\oplus b\oplus c=xy\oplus yz\oplus xz|x,y,z)$ on the left-hand side of Equation (\ref{cca}) is higher than the maximum Svetlichny local correlation strength. This is because $P(a\oplus b\oplus c=xy\oplus yz\oplus xz|x,y,z)$ can lead to the violation of SI, and if it doesn't, the legitimate users would abort the key distribution. Therefore, in order to maximize the value of $q_{L}$, Eve should utilize the local correlation $P^{L}(a\oplus b\oplus c=xy\oplus yz\oplus xz|x,y,z)$ as the maximal Svetlichny local correlation, while employing the nonlocal correlation $P^{NL}(a\oplus b\oplus c=xy\oplus yz\oplus xz|x,y,z)$ corresponding to the maximum quantum Svetlichny nonlocal correlation (GHZ correlation). When Eve is mimicking the correlation $P(a\oplus b\oplus c=xy\oplus yz\oplus xz|x,y,z)$ in Eq. (\ref{si3}), we can derive the following equation for $q_{L}$:
\begin{eqnarray}
\frac{3}{4} q_{L}+\left(\frac{1}{2}+\frac{\sqrt{2}}{4}\right)(1-q_{L})=(2p-1)^3 \left(\frac{1}{2}+\frac{\sqrt{2}}{4}\right)+\frac{1-(2p-1)^3}{2}.
\label{eq-ql}
\end{eqnarray}
By solving this equation, we obtain the expression of $q_{L}$
\begin{eqnarray}
q_{L}=\frac{2(\sqrt{2}-3\sqrt{2}p+6\sqrt{2}p^2-4\sqrt{2}p^3)}{\sqrt{2}-1}.
\label{ql}
\end{eqnarray}
We observe that when the measurement is perfect, represented by $p=1$, the value of $q_{L}$ becomes zero. This implies that Eve has no opportunity to eavesdrop in this particular scenario.

We will now utilize the Devetak-Winter formula \cite{Devetak2005} to calculate the extractable secret-key rate:
\begin{eqnarray}
r_{DW}\geq H(A|E)-H(A|B,C).
\label{key rate}
\end{eqnarray}
In this equation, $A$, $B$, $C$, and $E$ represent the measurement results of Alice, Bob, Carol, and Eve, respectively. For the sake of convenience, we assume these variables take values of $-1$ or $1$, as in Eq. (\ref{si}). $H(A|E)$ and $H(A|B,C)$ are both conditional Shannon entropies. $H(A|E)$ quantifies the correlation between Alice and Eve, while $H(A|B,C)$ quantifies the correlation between Alice and Bob and Carol.
It should be emphasized that Eq. (\ref{key rate}) deviates slightly from its standard form for the two-party case due to the presence of collective correlations in three-party key strings, as opposed to the pairwise correlations in the two-party scenario.

We proceed to calculate $H(A|E)$. As previously mentioned, Eve has the capability to emulate the correlation $P(a\oplus b\oplus c=xy\oplus yz\oplus xz|x,y,z)$ that is observed by the legitimate users with a measurement accuracy of $p$, as described in Eq. (\ref{cca}). Here, $P^{L}(a\oplus b\oplus c=xy\oplus yz\oplus xz|x,y,z)$ represents the maximal Svetlichny local correlation that achieves the classical bound of $\frac{3}{4}$ for the SI in Eq. (\ref{si2}), while $P^{NL}(a\oplus b\oplus c=xy\oplus yz\oplus xz|x,y,z)$ represents the GHZ correlation that achieves the quantum bound of $\frac{1}{2}+\frac{\sqrt{2}}{4}$ for the same SI. Additionally, we have $q_{L}$ defined in Eq. (\ref{ql}).
Given that $P^{NL}$ represents the GHZ correlation generated by the maximum entanglement state with a proper measurement protocol, the perfect monogamy property ensures that Eve has no access to any information regarding Alice's outcome $A$ in this scenario. However, we make the assumption that Eve possesses the capability to obtain complete information about $A$ when local correlation $P^{L}$ is distributed. Hence, for every measurement result $E$ obtained by Eve, we can establish the conditional probabilities $P(A=E|E)=\frac{1+q_{L}}{2}$ and $P(A\neq E|E)=\frac{1-q_{L}}{2}$, leading to
\begin{eqnarray}
H(A|E)=-\sum_{A,E}{P(E) P(A|E)\log_{2}{P(A|E)}}=h\left(\frac{1+q_{L}}{2}\right).
\label{key rate 2}
\end{eqnarray}
Here, $h(x)=-x\log_{2}{x}-(1-x)\log_{2}{(1-x)}$ represents the binary entropy, and $q_{L}$ is defined in Eq. (\ref{ql}).

\begin{figure}[t]
\includegraphics[width=0.80\columnwidth,
height=0.50\columnwidth]{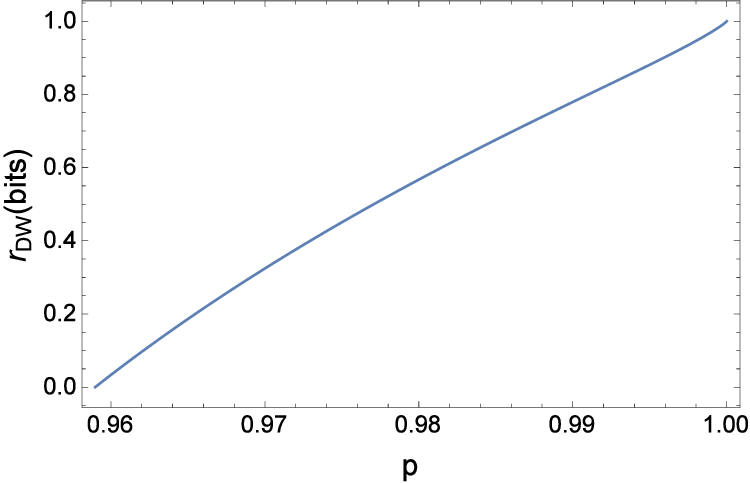} \caption{Extractable secret-key rate $r_{DW}$ against measurement accuracy $p$ under the condition of convex combination attack. Here, the $r_{DW}$ is given by Eq. (\ref{key rate 4}) . When $p>0.958968$ the $r_{DW}$ is greater than zero.
The value of $r_{DW}$ monotonously approaches $1$ as the value of measurement accuracy $p$ increases, and $r_{DW}=1$ when $p=1$.}
\label{fig1}
\end{figure}

Let's move on to calculate $H(A|B,C)$. As mentioned earlier, the legitimate users extract raw keys from the outcomes of the measurements corresponding to the settings $(0,0,0)$, $(0,\frac{\pi}{2},\frac{\pi}{2})$, $(\frac{\pi}{2},0,\frac{\pi}{2})$, and $(\frac{\pi}{2},\frac{\pi}{2},0)$.
Assuming that the probability of choosing each measurement is equal among the legitimate users, it is evident that the calculated $H(A|B,C)$ would be the same for each measurement setting. Therefore, we will focus on calculating the case of the measurement setting $(0,0,0)$.
In this case, all three legitimate users perform the $x$-direction spin measurement, denoted as $\sigma_{x}$. By applying $\sigma_{x}\otimes\sigma_{x}\otimes\sigma_{x}\ket{\Psi}_{GHZ}=\ket{\Psi}_{GHZ}$, we can conclude that if the measurement is perfect, we will obtain $P(A=BC|B,C)=1$ for every measurement result combination $B$ and $C$ obtained by Bob and Carol.
For the scenario where the measurement accuracy is $p$, calculating $P(A=BC|B,C)$ for any combination $B$ and $C$ is straightforward. We find that $P(A=BC|B,C)=p^3+3p(1-p)^2$. Therefore, the expression for $H(A|B,C)$ is obtained as follows:
\begin{eqnarray}
H(A|B,C)&=&-\sum_{A,B,C}{P(B,C)P(A|B,C)\log_{2}{P(A|B,C)}}\nonumber\\
&=&h(p^3+3p(1-p)^2).
\label{key rate 3}
\end{eqnarray}

Substituting the results of Eq.(\ref{key rate 2}) and Eq.(\ref{key rate 3}) into Eq.(\ref{key rate}), we can obtain
\begin{eqnarray}
r_{DW}\geq h\left(\frac{1+q_{L}}{2}\right)-h(p^3+3p(1-p)^2).
\label{key rate 4}
\end{eqnarray}
From Eq. (\ref{ql}) and Eq. (\ref{key rate 4}), we can conclude that the extractable secret-key rate is positive only when $p>0.958968$. We refer to this threshold measurement accuracy as $p_{th}$. It is worth noting that $p_{th}>p_{cr}$. Furthermore, we observe that this property holds for the general case of $n$-party, as we will discuss later.
In Fig. \ref{fig1}, we depict the extractable secret-key rate $r_{DW}$, which demonstrates that as the measurement accuracy $p$ increases, $r_{DW}$ approaches $1$ monotonically. When $p=1$, we have $r_{DW}=1$.

\section{$N$-party key distribution with imperfect measurement accuracy}

The same method can be directly applied to the scenario of $n$-party case. To study the impact of measurement accuracy, it is convenient to use the SI of $n$-party in the form of probabilities:
\begin{eqnarray}
\frac{1}{2^{n}}\sum_{\{x_{i}\}}P\Big(\sum_{i}^{n}a_{i}=\sum_{i<j\leq n}x_{i}x_{j}|x_{1},x_{2},...,x_{n}\Big)\leq \frac{3}{4},
\label{nsi}
\end{eqnarray}
Here, the $x_{i}$ values represent the measurement choices of the legitimate users and can be either $0$ or $1$. The $a_{i}$ values represent the corresponding measurement outcomes and also take on values of $0$ or $1$. The notation $\{x_{i}\}$ represents an $n$-tuple $x_{1},…,x_{n}$. The summation symbols $\sum_{i}^{n}$ and $\sum_{i<j\leq n}$ both denote summation modula $2$. $P$ is the probability that $\sum_{i}^{n}a_{i}=\sum_{i<j\leq n}x_{i}x_{j}$ with given $x_{1},x_{2},...,x_{n}$.

Similar to the approach for the three-party case, we assume that in the scenario of perfect measurements, the probability distribution of the measurement results is denoted as $P^{\prime}\Big(\sum_{i}^{n}a_{i}=\sum_{i<j\leq n}x_{i}x_{j}|x_{1},x_{2},...,x_{n}\Big)$. Since the legitimate users share an $n$-particle GHZ state, $P^{\prime}\Big(\sum_{i}^{n}a_{i}=\sum_{i<j\leq n}x_{i}x_{j}|x_{1},x_{2},...,x_{n}\Big)$ can achieve the quantum bound of $n$-party SI with a suitable measurement protocol. It is expressed as follows
\begin{eqnarray}
\frac{1}{2^{n}}\sum_{\{x_{i}\}}P^{\prime}\Big(\sum_{i}^{n}a_{i}=\sum_{i<j\leq n}x_{i}x_{j}|x_{1},x_{2},...,x_{n}\Big)= \frac{1}{2}+\frac{\sqrt{2}}{4}.
\label{nsi-qbound}
\end{eqnarray}
If we consider a measurement accuracy of $p$, the probability distribution of the measurement results can be represented as $P\Big(\sum_{i}^{n}a_{i}=\sum_{i<j\leq n}x_{i}x_{j}|x_{1},x_{2},...,x_{n}\Big)$. The relationship between $P\Big(\sum_{i}^{n}a_{i}=\sum_{i<j\leq n}x_{i}x_{j}|x_{1},x_{2},...,x_{n}\Big)$ and $P^{\prime}\Big(\sum_{i}^{n}a_{i}=\sum_{i<j\leq n}x_{i}x_{j}|x_{1},x_{2},...,x_{n}\Big)$ is given by
\begin{eqnarray}
P\Big(\sum_{i}^{n}a_{i}=\sum_{i<j\leq n}x_{i}x_{j}|x_{1},x_{2},...,x_{n}\Big)&=&(2p-1)^n P^{\prime}\Big(\sum_{i}^{n}a_{i}=\sum_{i<j\leq n}x_{i}x_{j}|x_{1},x_{2},...,x_{n}\Big)\nonumber\\
&&+\frac{1-(2p-1)^n}{2}.
\label{nsi2}
\end{eqnarray}
From Eq. (\ref{nsi-qbound}) and Eq. (\ref{nsi2}), we obtain
\begin{eqnarray}
\frac{1}{2^{n}}\sum_{\{x_{i}\}}P\Big(\sum_{i}^{n}a_{i}=\sum_{i<j\leq n}x_{i}x_{j}|x_{1},x_{2},...,x_{n}\Big)=
(2p-1)^n \left(\frac{1}{2}+\frac{\sqrt{2}}{4}\right)+\frac{1-(2p-1)^n}{2}.
\label{nsi3}
\end{eqnarray}
By using Eq. (\ref{nsi3}), we can calculate the critical measurement accuracy $p_{cr}$ needed to violate $n$-party SI.

%%%%%%%%%%%%%%%%%%%%%%%%%%%%%%%%%%%%%%%%%%%%%%%%%%%%%%%%%%%%%%%%%%%%%%%%%%%%%%%%
\begin{figure}[t]
\includegraphics[width=0.80\columnwidth,
height=0.50\columnwidth]{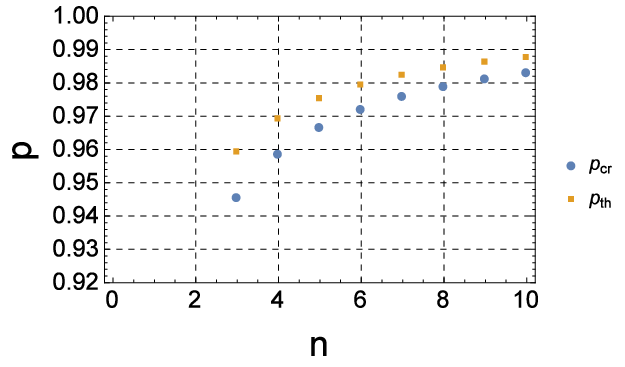} \caption{We present the graphical representation of $p_{th}$ and $p_{cr}$ for $n$-party scenarios ranging from $n=3$ to $n=10$. Notably, it is evident that in all cases, $p_{th}$ surpasses $p_{cr}$. As the value of $n$ increases, both $p_{th}$ and $p_{cr}$ exhibit a monotonically and rapidly approaching trend towards $1$.}
\label{fig2}
\end{figure}

\begin{figure}[t]
\includegraphics[width=0.80\columnwidth,
height=0.50\columnwidth]{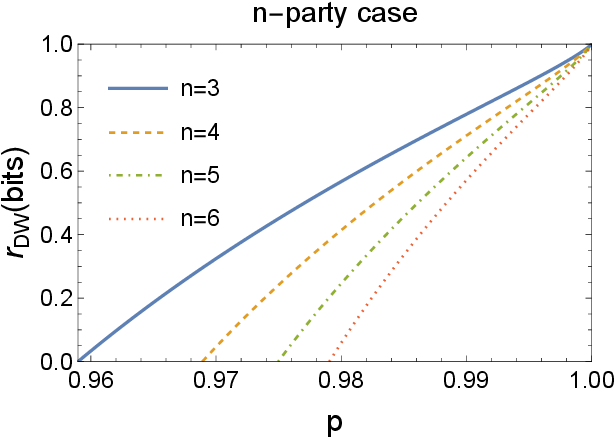} \caption{Extractable secret-key rates $r_{DW}$ for the $3$-party, $4$-party, $5$-party, and $6$-party cases can be computed by using Eq. (\ref{n key rate 2}), under the condition of convex combination attack. In all cases, the value of $r_{DW}$ monotonously approaches $1$ as the measurement accuracy $p$ increases, and when $p=1$, $r_{DW}$ reaches its maximum value of $1$.}
\label{fig3}
\end{figure}

%%%%%%%%%%%%%%%%%%%%%%%%%%%%%%%%%%%%%%%%%%%%%%%%%%%%%%%%%%%%%%%%%%%%%%%%%%%%%%%%%%%%%

We continue to consider the scenario where Eve employs a convex combination attack. She mimic the correlation observed by the legitimate users, which can be expressed by the following equation
\begin{eqnarray}
P\Big(\sum_{i}^{n}a_{i}=\sum_{i<j\leq n}x_{i}x_{j}|x_{1},x_{2},...,x_{n}\Big)&=&q_{L} P^{L}\Big(\sum_{i}^{n}a_{i}=\sum_{i<j\leq n}x_{i}x_{j}|x_{1},x_{2},...,x_{n}\Big)\nonumber\\
&+&(1-q_{L})P^{NL}\Big(\sum_{i}^{n}a_{i}=\sum_{i<j\leq n}x_{i}x_{j}|x_{1},x_{2},...,x_{n}\Big).\nonumber\\
\label{ncca}
\end{eqnarray}
In order to maximize the value of $q_{L}$, Eve should use the local correlation $P^{L}$ as the maximal $n$-party Svetlichny local correlation.
Additionally, she should employ the nonlocal correlation $P^{NL}$ corresponding to the maximum quantum $n$-party Svetlichny nonlocal correlation, which is the $n$-qubit GHZ correlation. When Eve is mimicking the correlation $P\Big(\sum_{i}^{n}a_{i}=\sum_{i<j\leq n}x_{i}x_{j}|x_{1},x_{2},...,x_{n}\Big)$ in Eq. (\ref{nsi3}), we can derive the following equation for $q_{L}$
\begin{eqnarray}
\frac{3}{4} q_{L}+\left(\frac{1}{2}+\frac{\sqrt{2}}{4}\right)(1-q_{L})=(2p-1)^n \left(\frac{1}{2}+\frac{\sqrt{2}}{4}\right)+\frac{1-(2p-1)^n}{2}.
\label{neq-ql}
\end{eqnarray}
By solving this equation, we obtain the expression for $q_{L}$
\begin{eqnarray}
q_{L}=\frac{\sqrt{2}\left[1-(2p-1)^n\right]}{\sqrt{2}-1}.
\label{nql}
\end{eqnarray}
We also observe that when the measurement is perfect, this is represented by $p=1$, the value of $q_{L}$ becomes zero. This implies that in a perfect measurement scenario, Eve has no opportunity to eavesdrop.

We finally utilize the Devetak-Winter formula to calculate the extractable secret-key rate of the $n$-party case:
\begin{eqnarray}
r_{DW}\geq H(A_{1}|E)-H(A_{1}|A_{2},A_{3},...,A_{n}).
\label{n key rate}
\end{eqnarray}
In this equation, $E$ still represent the measurement result of Eve, and $A_{i}$'s represent the measurement results of the legitimate users. For the sake of convenience, we assume these variables take values of $-1$ or $1$. $H(A_{1}|E)$ and $H(A_{1}|A_{2},A_{3},...,A_{n})$ are both conditional Shannon entropies. $H(A_{1}|E)$ quantifies the correlation between Alice and Eve, while $H(A_{1}|A_{2},A_{3},...,A_{n})$ quantifies the correlation between Alice and the other legitimate users. Without going into the calculation process, we will directly present the result
\begin{eqnarray}
r_{DW}\geq h\left(\frac{1+q_{L}}{2}\right)-h\left(\frac{1+(2p-1)^n}{2}\right),
\label{n key rate 2}
\end{eqnarray}
where $q_{L}$ is defined in Eq. (\ref{nql}).

By using Eq. (\ref{nql}) and Eq. (\ref{n key rate 2}), we can calculate the threshold measurement accuracy $p_{th}$ required to achieve a positive extractable secret-key rate for any $n$-party scenario. Interestingly, we observe that in all cases, $p_{th}$ exceeds $p_{cr}$. In Fig. \ref{fig2}, we illustrate $p_{th}$ and $p_{cr}$ for $n$-party scenarios ranging from $n=3$ to $n=10$.
We can also compute the extractable secret-key rate $r_{DW}$ using Eq. (\ref{nql}) and Eq. (\ref{n key rate 2}). In Fig. \ref{fig3},  we present the extractable secret-key rates $r_{DW}$ for the $3$-party, $4$-party, $5$-party, and $6$-party scenarios. It can be observed that in all cases, the value of $r_{DW}$ monotonously approaches $1$ as the measurement accuracy $p$ increases. When $p=1$, $r_{DW}$ reaches its maximum value of $1$.

\section{Werner state with imperfect measurement accuracy}

\begin{figure}[htbp]
\includegraphics[width=0.80\columnwidth,
height=0.50\columnwidth]{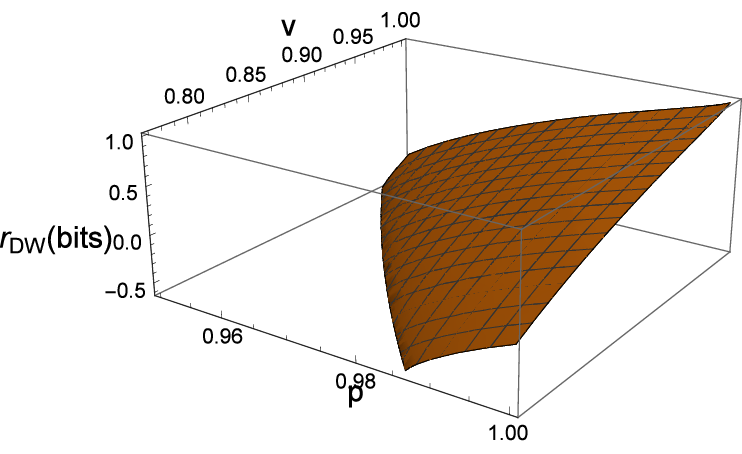} \caption{The extractable secret-key rates $r_{DW}$ in the scenario where the legitimate users have imperfect measurement accuracy and share a Werner state can be described as a function of the visibility $v$ and the measurement accuracy $p$.
In this case, $r_{DW}$ will only be greater than $0$ if both the visibility $v$ and the measurement accuracy $p$ are high. }
\label{fig4}
\end{figure}

\begin{figure}[htbp]
\includegraphics[width=0.80\columnwidth,
height=0.50\columnwidth]{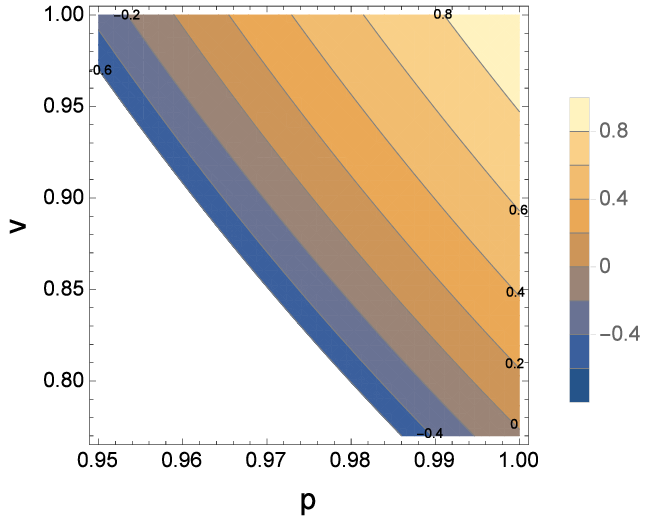} \caption{The contour map depicts the extractable secret-key rate $r_{DW}$, as shown in Fig. \ref{fig4}. It clearly illustrates the specific region in the $v$-$p$ plane where $r_{DW}$ exceeds zero.}
\label{fig5}
\end{figure}

Quantum entanglement is a fragile resource that is often compromised by noise during transmission. In this section, we consider a scenario where the targeted GHZ state is subjected to white noise during transmission and is transformed into the Werner state. We will calculate the extractable secret-key rate in this scenario, taking into account the fact that the legitimate users have imperfect measurement accuracy.
Specifically, we will focus on the three-party case, but the methodology can be generalized to the more general $n$-party case.
The three-qubit Werner state is \cite{PhysRevA.40.4277}
\begin{eqnarray}
\rho^{v}=v\ket{\Psi}_{GHZ}\bra{\Psi}+(1-v)\frac{\mathbb{I}}{8},
\label{werner}
\end{eqnarray}
we call $v\in[0,1]$ the visibility, it quantify the degree to which the entanglement is maintained.
If the three legitimate users share this Werner state but have perfect measurement accuracy, they can apply a suitable measurement protocol to observe the correlation $P^{\prime}(a\oplus b\oplus c=xy\oplus yz\oplus xz|x,y,z)$, which satisfy
\begin{eqnarray}
\frac{1}{8}\sum_{x,y,z}P^{\prime}(a\oplus b\oplus c=xy\oplus yz\oplus xz|x,y,z)=v \left(\frac{1}{2}+\frac{\sqrt{2}}{4}\right)+(1-v)\frac{1}{2}
\label{wp}
\end{eqnarray}
If the measurement accuracy is $p$, we denote the observed correlation as $P(a\oplus b\oplus c=xy\oplus yz\oplus xz|x,y,z)$. The relationship between this correlation and $P^{\prime}(a\oplus b\oplus c=xy\oplus yz\oplus xz|x,y,z)$ has been expressed in Eq. (\ref{si3}). By utilizing Eq.(\ref{wp}), we can derive
\begin{eqnarray}
\frac{1}{8}\sum_{x,y,z}P(a\oplus b\oplus c=xy\oplus yz\oplus xz|x,y,z)&=&(2p-1)^3\left[v \left(\frac{1}{2}+\frac{\sqrt{2}}{4}\right)+(1-v)\frac{1}{2}\right]\nonumber\\
&&+\frac{1-(2p-1)^3}{2}.
\label{wp2}
\end{eqnarray}

In order to maximize the value of $q_{L}$, Eve should utilize the maximal Svetlichny local correlation $P^{L}$ and the maximum quantum Svetlichny nonlocal correlation $P^{GHZ}$ (generated from the GHZ state) in her convex combination attack. This will allow her to mimic the correlation $P(a\oplus b\oplus c=xy\oplus yz\oplus xz|x,y,z)$ in Eq. (\ref{wp2}) as
\begin{eqnarray}
P(a\oplus b\oplus c=xy\oplus yz\oplus xz|x,y,z)=q_{L} P^{L}+(1-q_{L}) P^{GHZ}.
\label{wcca}
\end{eqnarray}
Combining Eq. (\ref{wp2}) and Eq. (\ref{wcca}), we obtain the following equation for $q_{L}$
\begin{eqnarray}
\frac{3}{4} q_{L}+\left(\frac{1}{2}+\frac{\sqrt{2}}{4}\right)(1-q_{L})&=&(2p-1)^3\left[v \left(\frac{1}{2}+\frac{\sqrt{2}}{4}\right)+(1-v)\frac{1}{2}\right]\nonumber\\
&&+\frac{1-(2p-1)^3}{2}.
\label{w-eq-ql}
\end{eqnarray}
By solving the above equation, we obtain the expression for $q_{L}$
\begin{eqnarray}
q_{L}=(2+\sqrt{2})\left[1-(2p-1)^3 v\right].
\label{w-ql}
\end{eqnarray}

We can now calculate the extractable secret-key rate by using Eq. (\ref{key rate}).
In the scenario of the Werner state, we still have $H(A|E)=h\left(\frac{1+q_{L}}{2}\right)$. However, the expression for $q_{L}$ in this case is given by Eq. (\ref{w-ql}). And then we find $P(A=BC|B,C)=\left[\frac{1+(2p-1)^3}{2}\right] v+\frac{1}{2} (1-v)$ in this case, so $H(A|B,C)=h\left(\left[\frac{1+(2p-1)^3}{2}\right] v+\frac{1}{2} (1-v)\right)$. Finally, we obtain
\begin{eqnarray}
r_{DW}\geq h\left(\frac{1+q_{L}}{2}\right)-h\left(\left[\frac{1+(2p-1)^3}{2}\right] v+\frac{1}{2} (1-v)\right).
\label{w-key rate}
\end{eqnarray}
So in the scenario where legitimate users have imperfect measurement accuracy and share a Werner state, the extractable secret-key rate $r_{DW}$ (Eq. (\ref{w-key rate})) depends on the visibility $v$ and the measurement accuracy $p$, we illustrate it in Fig. \ref{fig4}. For $r_{DW}$ to be greater than zero, it is necessary for both the visibility $v$ and the measurement accuracy $p$ to be at high levels. Fig. \ref{fig5} shows the contour map of $r_{DW}$, clearly illustrating the specific region in the $v$-$p$ plane where $r_{DW}$ is greater than zero.

\section{Conclusion}

The imperfection of measurements in real-world scenarios can compromise the performance of DIQKD protocols. In this study, we investigate the impact of measurement imperfections on our multi-party DIQKD protocol \cite{Xiang2023}, which relies on the violation of the SI, assuming an eavesdropper employing the convex combination attack. We analyze both the three-party DIQKD case and the general $n$-party scenario. Our findings reveal that the threshold measurement accuracy $p_{th}$ required to achieve positive extractable secret-key rates is always higher than the critical measurement accuracy $p_{cr}$ needed to violate the corresponding SI. We present a graphical representation of these thresholds for $n$-party scenarios ranging from $n=3$ to $n=10$. The graph demonstrates that as the value of $n$ increases, both $p_{th}$ and $p_{cr}$ show a trend of approaching unity monotonically and rapidly.
We additionally establish the relationship between the measurement accuracy $p$ and the extractable secret-key rate $r_{DW}$ in the general $n$-party scenario, considering an eavesdropper utilizing the convex combination attack. Our findings demonstrate that as the measurement accuracy $p$ increases, the value of $r_{DW}$ steadily approaches $1$. Notably, when $p=1$, $r_{DW}$ reaches its maximum value of $1$. Furthermore, we explore a combined scenario that encompasses a non-maximally entangled state combined with imperfect measurements. Specifically, we assume that the initial GHZ state emitted is subject to noise during transmission, resulting in a Werner state. Within this framework, we calculate and illustrate the relationship between the extractable secret-key rate $r_{DW}$, the visibility $v$ of the Werner state, and the measurement accuracy $p$, specifically focusing on the three-party scenario.

Overall, this study provides valuable insights into the impact of imperfect measurement accuracy on the security and performance of multi-party DIQKD protocols. It highlights the significance of high measurement accuracy in achieving positive secret-key rates and maintaining the violation of the SI.
It should be emphasized that the threshold value $p_{th}$ calculated in this paper represents only a lower bound for the required measurement accuracy to extract a positive key. This is because the key rate is calculated assuming that Eve utilizes the convex combination attack. However, it is important to note that this specific attack strategy is not guaranteed to be optimal for her. As a result, the actual key rate may be lower than the values presented in this paper, thereby increasing the threshold value for the measurement accuracy required to achieve positive key rates.

\section*{CRediT authorship contribution statement}
~\textbf{Yang Xiang}: Conceptualization, Methodology, Formal analysis, Investigation, Writing-original draft, Writing-review and editing.

\section*{Declaration of competing interest}
~The authors declare that they have no known competing financial interests or personal relationships that could have appeared to influence the work reported in this paper.

\section*{Data availability}
~Data will be made available on request.

\section*{Acknowledgments}
~This work is supported by the National Natural Science Foundation of China under Grant No. 11005031.

%\section*{Data Availability Statement}
%~This manuscript has no associated data or the data will not be deposited. [Author’s
%comments: All relevant data are in the paper itself.]

%\vskip 0.5 cm

%{\em Acknoledgments.---}

%\begin{thebibliography}{100}
%\bibitem{panch} S.Pancharatnam, Proc.Indian Acad. Sci. A
%{\bf44}, 247(1956).

\bibliographystyle{apsrev4-1}
\bibliography{mqkd}

\end{document}